\documentclass[fleqn,twoside]{article}
\usepackage{espcrc2,graphicx}

\usepackage{graphicx}
\usepackage[figuresright]{rotating}

\def\d {\mathrm{d}}

\def\la{\mathrel{\mathchoice {\vcenter{\offinterlineskip\halign{\hfil
$\displaystyle##$\hfil\cr<\cr\sim\cr}}}
{\vcenter{\offinterlineskip\halign{\hfil$\textstyle##$\hfil\cr
<\cr\sim\cr}}}
{\vcenter{\offinterlineskip\halign{\hfil$\scriptstyle##$\hfil\cr
<\cr\sim\cr}}}
{\vcenter{\offinterlineskip\halign{\hfil$\scriptscriptstyle##$\hfil\cr
<\cr\sim\cr}}}}}

\newcommand{\AmS}{{\protect\the\textfont2
  A\kern-.1667em\lower.5ex\hbox{M}\kern-.125emS}}

\title{GZK horizon and magnetic fields}

\author{Etienne Parizot\\
\addressmark{Institut de Physique Nucl\'eaire d'Orsay, IN2P3-CNRS/Universit\'e Paris-Sud, \\ 15, rue Georges Cl\'emenceau, 91406 Orsay Cedex, France}}
       
\begin{document}

\begin{abstract}
We discuss some aspects of the propagation of high-energy cosmic rays (CRs) in turbulent magnetic fields, and propose a formula for the diffusion coefficient based on accurate simulations in a wide energy range. We discuss the transition between ballistic and diffusive regimes and the magnetic horizon of CR sources, showing how magnetic fields of a few nG could modify the shape of the GZK feature. Such fields would roughly be in equipartition with the extragalactic CRs, and could build through the resonant growth of waves within the lifetime of the universe.
\end{abstract}

\maketitle

\section{Introduction}

Two independent problems must be distinguished in the physics and phenomenology of ultra-high-energy cosmic rays (UHECRs). The first one relates to their production and involves the identification of efficient astrophysical or exotic sources. The main issues are the prediction of a (sufficiently high) maximum energy, a definite source spectrum and source composition, and global energetics. Most models do not currently have such a wide predictive power. The main challenge for astrophysical models appears to be the acceleration of particles up to energies around $3\,10^{20}$~eV, while the so-called top-down models encounter some difficulty to ``hide'' other types of radiation associated with the production and propagation of UHECRs.

The second, independent problem relates to UHECRs propagation in the universe, which involves two complementary aspects: energy losses (mostly via interactions with the CMB) and particle deflection in ambient magnetic fields. In the case of nuclei, a third issue is related to composition changes, as photo-nuclear interactions lead to the ``erosion'' of primary heavy nuclei.

As described forty years ago by Greisen\cite{Greisen66} and by Zatsepin and Kuzmin\cite{ZatKuz66} (GZK), a sudden decrease in the observed flux of cosmic-rays is expected when the energy of the CMB photons in the rest frame of the propagating proton (or nucleus) reaches values in excess of the pion photoproduction (or photodisintegration) threshold. This should occur around $8\,10^{19}$~eV. Whether such a ``cutoff'' has been observed or not is still unclear, as the two main experiments which could address this question, AGASA and HiRes, do not agree on the answer, although with a statistical significance of only $\sim 2\sigma$ \cite{DeMarco+03}. The determination of the CR spectrum above $\sim 10^{20}$~eV is one of the major goals of the Pierre Auger Observatory \cite{PAO}, which should provide its first results soon.

\section{UHECRs and magnetic fields}

In the original GZK argument and in most subsequent discussions, the extragalactic magnetic field has been assumed to be very low, say $B\la 10^{-11}$~G, so that its effect on particle trajectories could be neglected. Such an assumption is likely to be correct, indeed, as no mechanism has been proposed to produce higher magnetic fields over very large scales. Current data suggest that, except inside galaxies and possibly in the internal parts of galaxy clusters -- which represent a negligible fraction of the volume of the universe -- magnetic fields are generally well below the equipartition values which would be required to produce significant deflections. However, it is expected that the sources of UHECRs are located in high magnetic field regions, because virtually all energetic phenomena take place in strongly magnetized regions, and intense magnetic fields are certainly required for particle acceleration. Therefore, even if they are globally negligible in the universe, magnetic fields may play a role in the phenomenology of UHECRs, not only at acceleration, but also during their transport, and it is interesting to ask how they modify the original GZK argument and in particular change the shape of the GZK feature at ultra high energy.

An important remark should also be made here. It is well known that CRs accelerated in supernova remnants by the diffusive shock acceleration mechanism produce themselves the MHD waves thanks to which they diffuse back and forth and cross the shock front many times. The growth of Alfv\'en waves by the so-called \emph{streaming instability} occurs whenever an anisotropic flux of CRs is found with a global streaming velocity larger than the Alfv\'en speed. Just as the CRs are scattered by resonant waves having a wavelength comparable with their gyroradius, the waves which are amplified by a CR stream have wavelengths of the order of their gyroradius in the underlying field. In a simplified approach, one may evaluate the growth rate of the waves (in the linear regime) by writing that the decrease of the CR momentum parallel to the stream must be compensated by the increase of the wave momentum.

During a scattering time $\tau_{\mathrm{s}}\simeq\lambda_{\mathrm{s}}/v$, the resonant CRs are deflected by $\sim 90^\circ$, which amounts to a momentum transfer of $\varepsilon_{\mathrm{CR}} v/c^2$ per unit volume, where $\varepsilon_{\mathrm{CR}}$ is the CR energy density. The wave energy and momentum densities are related by $u_{\mathrm{w}} = p_{\mathrm{w}}v_{\mathrm{A}}$, where $v_{\mathrm{A}} = B_{0}/\sqrt{4\pi\rho}$ is the Alfv\'en velocity in the regular field, $B_{0}$. Thus:
\begin{equation}
\frac{\d u_{\rm w}}{\d t} = \varepsilon_{\mathrm{CR}}\frac{v}{c^{2}} \frac{v_{\rm A}}{\tau_{\rm s}}.
\label{eq:duDt}
\end{equation}
If $B_{1}$ is the wave field, so that $u_{\mathrm{w}} = B_{1}^2/2\mu_{0}$, and $r_{\mathrm{g}} = p/qB_{0}$ is the gyroradius of the resonant particles, we have
\begin{equation}
\tau_{\mathrm{s}}\simeq \left(\frac{B_{0}}{B_{1}}\right)^2 \frac{r_{\mathrm{g}}}{v} = \frac{u_{0}}{u_{\mathrm{w}}} \frac{r_{\mathrm{g}}}{v}
\label{eq:tauS}
\end{equation}
and $\d u_{\rm w}/\d t = u_{\rm w}(v_{\mathrm{A}}/r_{\mathrm{g}})(\varepsilon_{\mathrm{CR}}/u_{0})$ (for $v\simeq c$). The wave growth timescale is thus:
\begin{equation}
\tau_{\mathrm{w}} \simeq\frac{r_{\mathrm{g}}}{v_{\mathrm{A}}} \frac{u_{0}}{\varepsilon_{\mathrm{CR}}} \simeq \frac{r_{\mathrm{g}}}{v_{\mathrm{A}}} \frac{B_{0}^2}{B_{\mathrm{eq}}^2},
\label{eq:tauW}
\end{equation}
where we have introduced the ``equilibrium'' magnetic field, $B_{\mathrm{eq}}$, which may be reached if the waves can grow up to a point when the magnetic energy density is similar to that of the cosmic rays (at a given resonant energy). Modeling the CR distribution function (in the energy range of interest) as $\Phi(E) \simeq 3\,10^{24}\,\mathrm{eV}^2\mathrm{m}^{-2}\mathrm{s}^{-1}\mathrm{sr}^{-1}$, we find $\varepsilon_{\mathrm{CR}}(E)\sim\varepsilon_{\mathrm{CR}}(\ge \hspace{-4pt}E)\simeq (1.3\,10^{-7} \,\mathrm{eV}\,\mathrm{cm}^{-3})\times E_{\mathrm{EeV}}^{-1}$, where $E_{\mathrm{EeV}}$ is in units of $10^{18}$~eV. Thence:
\begin{equation}
B_{\mathrm{eq}} \simeq 2.2\,\mathrm{nG}\, E_{\mathrm{EeV}}^{-1/2}.
\label{eq:BEq}
\end{equation}

It should be stressed that such a magnetic field is much larger than originally assumed for UHECR propagation, and also larger than what is obtained with current models of extragalactic magnetic field generation (e.g. \cite{Dolag+02,Sigl+04}). Therefore, if wave amplification by cosmic rays in the extragalactic medium is indeed an efficient process, it may be the main source of magnetic field in the universe (outside galaxies and clusters), which would be very interesting in itself and would also have important consequences on the global field structure (wave content and topology).

\begin{figure*}[t]
\hfill
\includegraphics[width=0.40\linewidth]{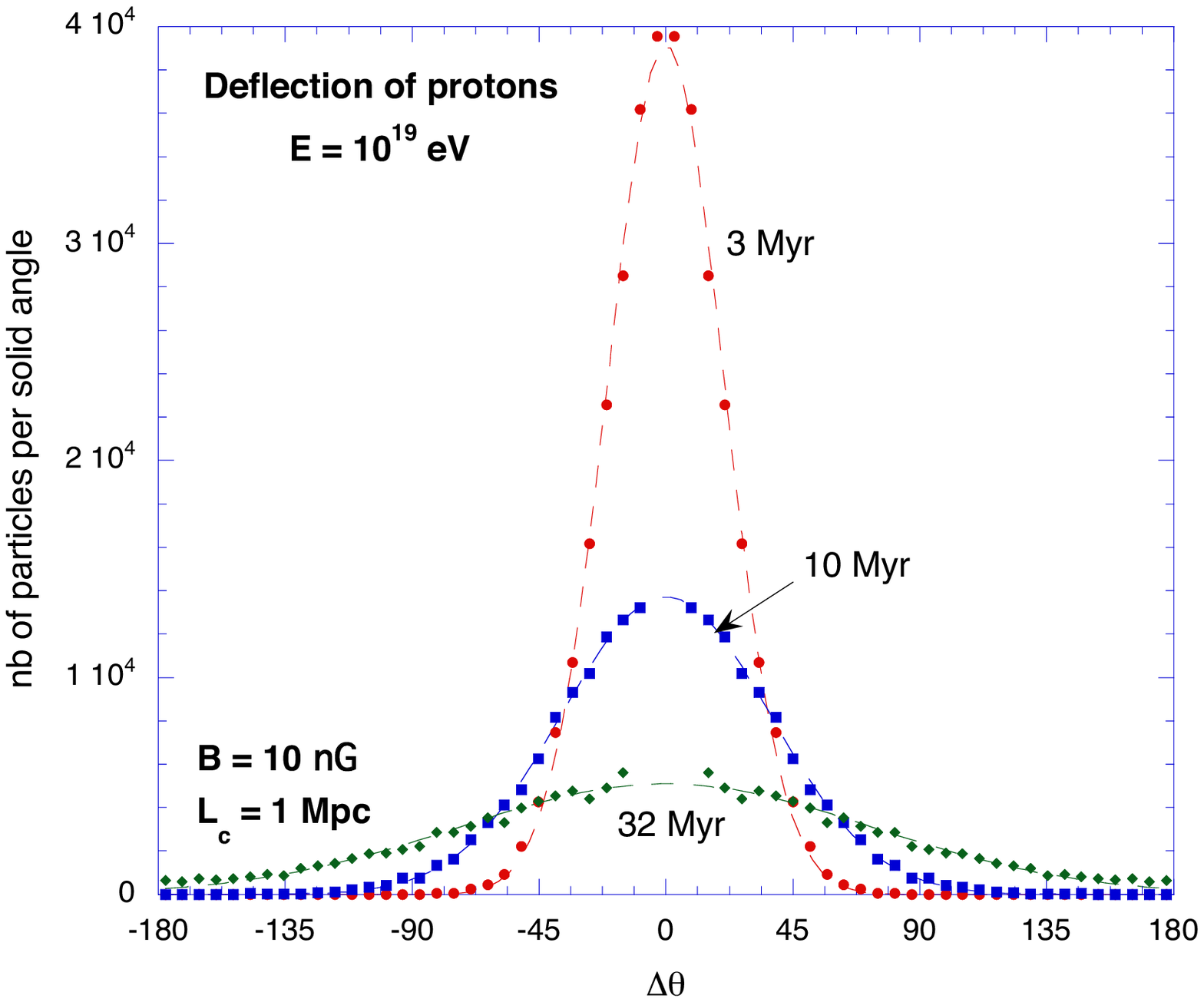}
\hfill
\includegraphics[width=0.40\linewidth]{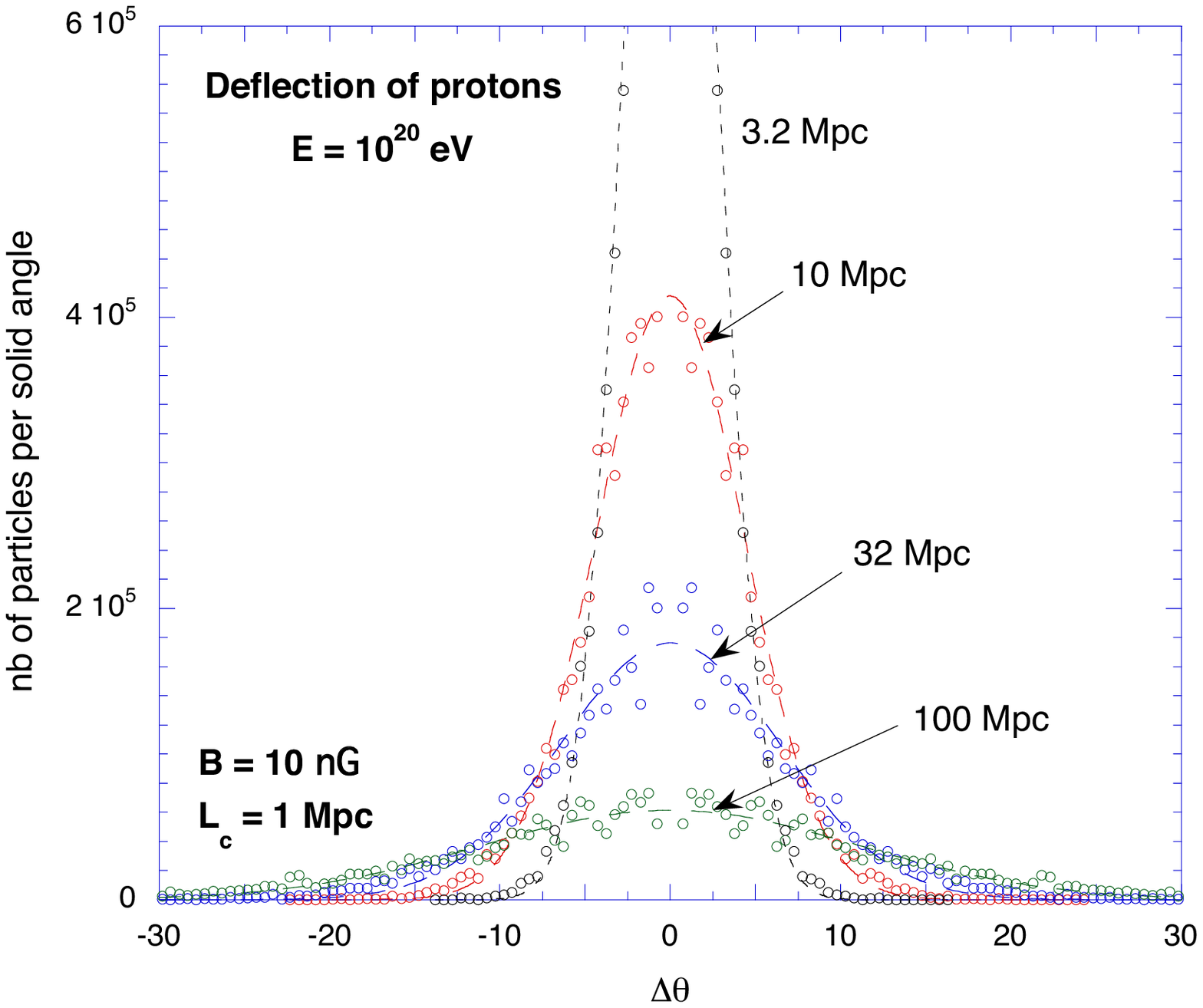}
\hfill
\vspace{-1cm}
\caption{Angular spread of protons of $10^{19}$~eV (left) and $10^{20}$~eV (right), propagating in a 10~nG turbulent magnetic field with $\lambda_{\mathrm{c}} = 1$~Mpc, after different times of flight, as indicated.}
\label{fig:protonDeflection}
\end{figure*}

Clearly, the detailed study of this process should involve an analysis of the wave growth away from the linear regime as well as the non-linear decay (through a turbulence-like cascade in wavenumbers) and the isotropization of the wave system (which is required since the particles are deflected by waves perpendicular to their stream, while they grow waves parallel to it). Such a study is postponed to a forthcoming work. We note, however, that in the above simplified approach, the growth timescale of the magnetic field at a scale resonant with extragalactic CRs is smaller than the age of the universe. Normalizing the extragalactic density, $n_{\mathrm{EG}}$, to the baryonic density of the universe, $n_{\mathrm{b}}$, such that $\Omega_{\mathrm{b}}h^2 = 0.02$, we have $v_{\mathrm{A}}\simeq 4\,10^5\,B_{\mathrm{nG}}(n_{\mathrm{EG}}/n_{\mathrm{b}})^{-1/2}$~cm/s. Besides,
\begin{equation}
r_{\mathrm{g}} \simeq 1.1\,\mathrm{Mpc} \times \frac{E_{\mathrm{EeV}}}{ZB_{\mathrm{nG}}}.
\label{eq:rG}
\end{equation}
For a resonant energy $E_{\mathrm{res}}$, Eq.~(\ref{eq:tauW}) then gives:
\begin{equation}
\tau_{\mathrm{w}}\simeq 5\,\mathrm{Gyr}\,\left(\frac{n_{\mathrm{EG}}}{10^{-2}\,n_{\mathrm{b}}}\right)^{1/2}\left(\frac{E_{\mathrm{res}}}{10^{18}\,\mathrm{eV}}\right)^{2}.
\label{eq:tauWNum}
\end{equation}
Note that $n_{\mathrm{EG}}$ may actually be much smaller than $10^{-2}n_{\mathrm{b}}$, because of the huge overdensity in the galaxy clusters, so that magnetic fields could indeed have been built, by now, by the CR population itself, up to scales resonant with very high energies. For completeness (although we stress once more that the above is a very crude approach), we give the resonant scale for a CR of energy E in the self-consistent field $B_{\mathrm{eq}}(E)$ (Eq.~\ref{eq:BEq}):
\begin{equation}
\lambda_{\mathrm{res}} \simeq 0.5\,\mathrm{Mpc} \left(\frac{E}{10^{18}\mathrm{eV}}\right)^{3/2}.
\label{eq:lambdaRes}
\end{equation}

\begin{figure*}[t]
\hfill
\includegraphics[width=0.47\linewidth]{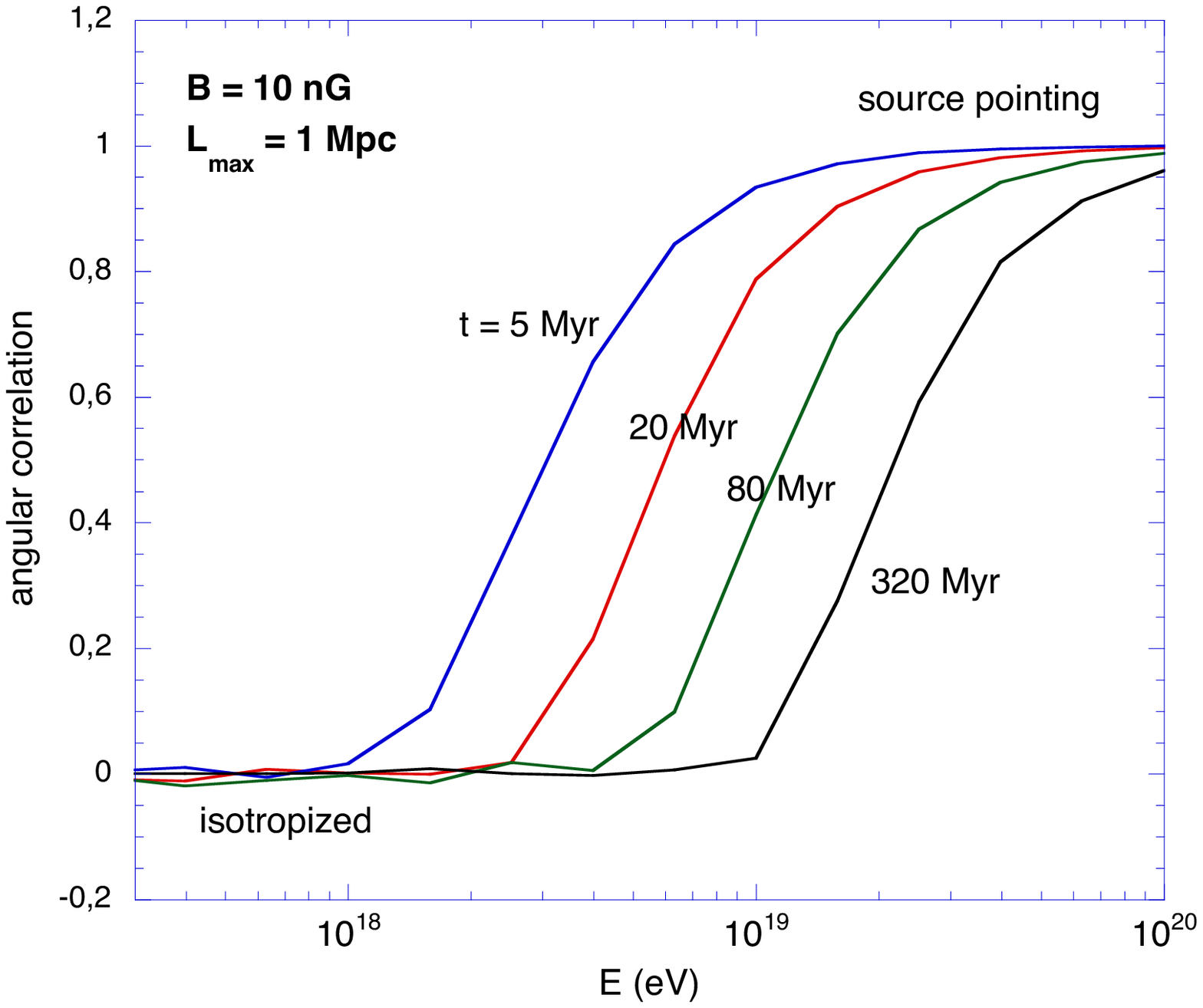}
\hfill
\includegraphics[width=0.47\linewidth]{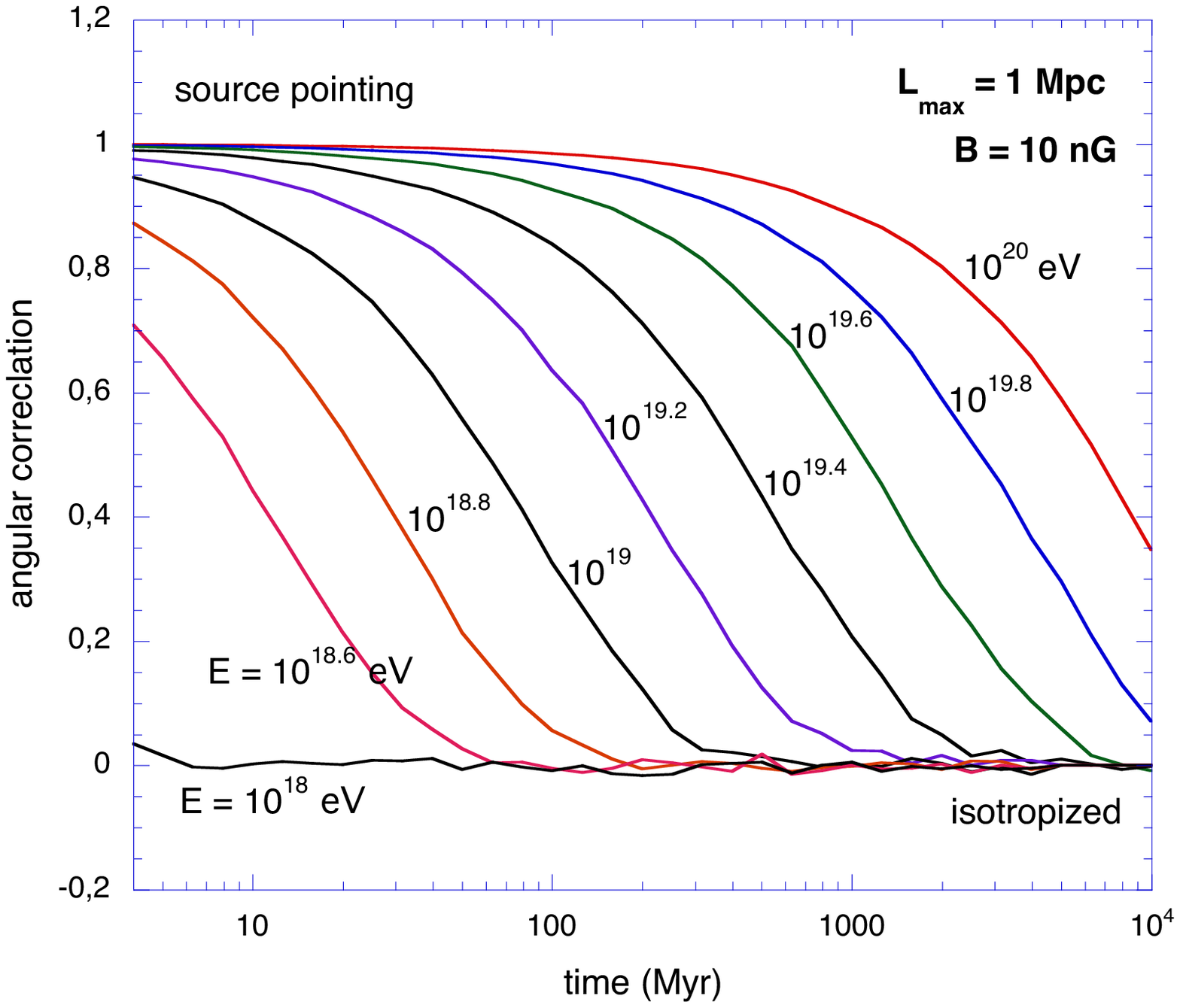}
\hfill
\vspace{-1cm}
\caption{Angular correlation, $\left<\cos\delta\theta\right>$, of a set of protons propagating in a 10~nG turbulent field, as a function of energy, at different times (left) and as a function of time, at different energies (right).}
\label{fig:angularCorrelation}
\end{figure*}

\section{Particle transport in turbulent fields}

Before turning to the possible influence of magnetic fields on the UHECR spectrum, we consider the general problem of CR transport in turbulent magnetic fields. In a galaxy like ours, the turbulent component is roughly as intense as the regular one (or even higher), with values around $5\mu$G, and a wavelength distribution compatible with a Kolmogorov spectrum, up to scales of 50--100~pc. In galaxy clusters, the observational data is very scarce, but $\delta B/B$ may also be of order 1, with coherence lengths of 10--100~kpc, over scales of a few Mpc. As for the deep extratragalactic space, current limits based on Faraday rotation measurements depend on the coherence length of the field, with possible values spanning a very wide range from totally negligible to a few tens of nG \cite{Kronberg94,Blasi+99}. In all the models considered, the \emph{regular} magnetic field in extragalactic voids is expected to be very low, and thus if magnetic fields are high enough to influence the UHECR spectrum, they should be dominated by the turbulent component.

The transport of charged particles in a turbulent field is conceptually very simple, since it comes down to integrating trajectories influenced by the sole Lorentz force. However, its detailed treatment is limited by the nature of the problem, which is essentially chaotic (and of course by our ignorance of the real configuration of the fields). We are thus forced to keep to a statistical description of the trajectories.

One should first note that individual trajectories only depend on the particle gyroradius (or energy per charge). In weak fields or at high energy (cf. Eq.~\ref{eq:rG}), when the particles have gyroradii much larger than the coherence length, $\lambda_{\mathrm{c}}$, they are only slightly deflected over that length, with $\delta\theta(\lambda_{\mathrm{c}})\simeq r_{g}/\lambda_{\mathrm{c}}$. Since the direction of the deflections are not correlated from one coherence length to the next, particles diffuse in angle (relative to the initial direction), with an angular spread $\sigma_{\theta}\sim t^{1/2}$. Figure~\ref{fig:protonDeflection} shows that Gaussian profiles fit very well the proton angular distributions at various times, as obtained with a numerical simulation. It also appears that the \emph{angular decorrelation} occurs on a longer timescale for higher energy particles, because of their larger rigidity. Figure~\ref{fig:angularCorrelation} also shows that the isotropization process spreads over typically one order of magnitude in energy, and two orders of magnitude in time.

Once the particles are completely isotropized, it is appropriate to abandon the description in terms of individual trajectories, and consider their propagation as a diffusion process, in which the average linear distance traveled by CRs increases as $\Delta r^2 = 6D\Delta t$ (this equation actually \emph{defines} the diffusion coefficient, $D$). To analyze in greater detail the transition from a \emph{ballistic} to a \emph{diffusive} propagation regime, we have performed a numerical simulation of the trajectory of charged particles in a purely turbulent field, represented by a sum of $N_{\mathrm{m}}$ modes as~\cite{GiaJok99}:
\begin{equation}
B = \sum_{1}^{N_{\mathrm{m}}}A_{k_{n}}\hat\xi_{n}\exp(ik_{n}z^\prime_{n} + i\beta_{n}),
\label{eq:BGC}
\end{equation}
where $\hat\xi_{n} = \cos\alpha_{n}\hat{\mathbf{x}}^\prime_{n} + i \sin\alpha_{n}\hat{\mathbf{y}}^\prime_{n}$, $\alpha_{n}$ and $\beta_{n}$ are random phases (chosen once for all) and $[x^\prime_{n},y^\prime_{n},z^\prime_{n}] = [\mathcal{R}(\theta_{n},\phi_{n})]\times[x,y,z]$ are coordinates obtained by a rotation of the reference frame bringing the z axis in the direction of the $n^\mathrm{th}$ contributing wave (i.e. $\mathbf{k}_{n}$ is in direction $[\theta_{n},\phi_{n}]$, also chosen randomly). The amplitude $A(k_{n})$ is determined as a function of $\|\mathbf{k}_{n}\|$ according to a specific assumption on the type of turbulence (here, we assume a Kolmogorov spectrum with wavelengths between $\lambda_{\mathrm{min}}$ and $\lambda_{\mathrm{max}}$).

\begin{figure}[t]
\hfill
\includegraphics[width=0.9\linewidth]{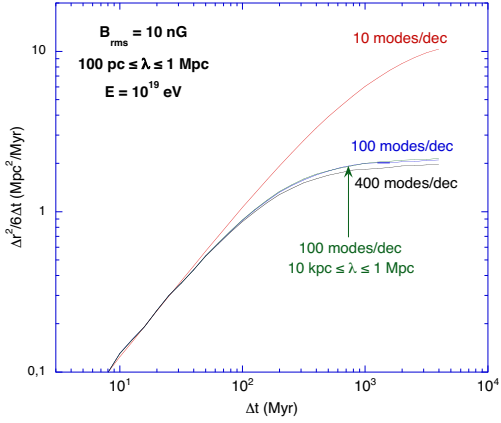}
\hfill
\vspace{-1cm}
\caption{Instantaneous effective diffusion coefficient (IEDC), $\Delta r^2/6\Delta t$, as a function of elapsed time, for a $10^{19}$~eV proton in a 10~nG field described with various numbers of modes per decade of wavelengths (in the indicated range).}
\label{fig:IEDCsModes}
\end{figure}

Then, in exactly the same spirit as in ref.~\cite{Casse+02}, we calculated the average quantity $\Delta r^2/6\Delta t$ -- which may be called the \emph{instantaneous effective diffusion coefficient} (IEDC) -- for 5000 trajectories, and plot the result as a function of elapsed time, $\Delta t$. Results are shown on Fig.~\ref{fig:IEDCsModes} for various levels of precision in the field description (numbers of modes per decade of wavenumbers, and ranges of wavelengths considered). The IEDC first increases linearly with time (as CRs propagate in straight line with velocity $c$), and then a transition occurs to a diffusive regime where the IEDC is constant, with a value which can be identified with the diffusion coefficient, $D(E)$, of the order of 2 Mpc/Myr in that case.

It is important to note that CR propagation cannot be described accurately without a sufficiently precise definition of the magnetic field. With only 10 modes per decade of wavenumbers, the transition towards diffusion occurs much later, and the obtained diffusion coefficient is much larger (both by an order of magnitude, see Fig.~\ref{fig:IEDCsModes}), than if one uses 100 modes per decade or more. Our calculations suggest that $\sim 100$ modes/dec offers a good compromise between accuracy and computation time. This casts some doubt on results obtained with propagation codes using a field described as constant over cubic cells of size $\lambda_{\mathrm{c}}$, as often found in the literature.

\begin{figure}[h]
\hfill
\includegraphics[width=0.97\linewidth]{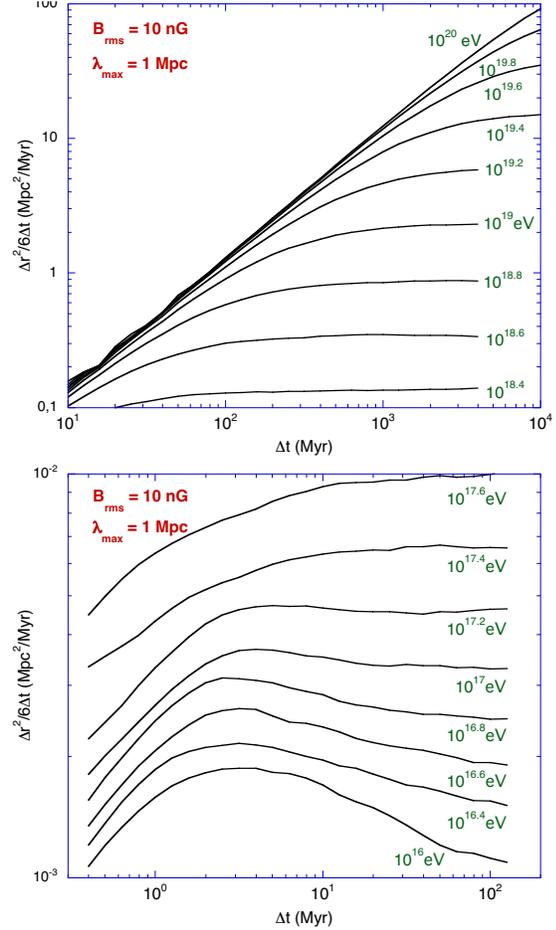}
\hfill
\vspace{-1cm}
\caption{IEDC for particles of various energies, from $10^{16}$ to $10^{20}$~eV, in a 10~nG field with a maximum wavelength $\lambda_{\mathrm{max}} = 1$~Mpc.}
\label{fig:IEDCs}
\end{figure}

Of course, the time at which the diffusion regime settles and the value of $D$ depend on the CR energy: lower energy particles enter earlier in the diffusion regime and have a smaller diffusion coefficient (see Fig.~\ref{fig:IEDCs} and next section). The diffusion regime can usually be modeled as a random walk at constant velocity (here $v = c$) with an isotropic redistribution of the direction after a ``mean free time'', which can be identified to the scattering time, $\tau_{\mathrm{s}}$, after which the direction of the particle is (on average) decorrelated from the initial direction (cf. Fig.~\ref{fig:angularCorrelation}). The diffusion coefficient can then be written as $D \simeq \frac{1}{3}c^2\tau_{\mathrm{s}} \simeq  \frac{1}{3}\lambda_{\mathrm{s}}c$, so that the energy dependences of $D$ and $\tau_{\mathrm{s}}$ are the same. Note that the standard Bohm diffusion coefficient is obtained for the smallest possible mean free path, $\lambda_{\mathrm{s}} = r_{\mathrm{g}}$: $D_{\mathrm{Bohm}}(E) = \frac{1}{3}cr_{\mathrm{g}}(E)$.

\section{Diffusion coefficients}

In Fig.~\ref{fig:IEDCs}, we show the evolution of IEDCs for different particle energies. A change of behaviour can be observed when the energy decreases. For the field parameters used here ($\lambda_{\mathrm{c}}=1$~Mpc, most probable value of $B = 10$~nG, or $\left<B\right> \simeq 11.3$~nG), a bump appears in the curves for particles of energy lower than $E_{0}\simeq 3\,10^{17}$~eV, corresponding to $r_{\mathrm{g}} \simeq \lambda_{\mathrm{c}}/2\pi$ ($\lambda_{\mathrm{c}} = \lambda_{\mathrm{max}}/5$ for a Kolmogorov spectrum\cite{Harari+02}). For energies larger than $E_{0}$, the diffusion process settles as described above, from coherent deflection to angular diffusion, isotropization and spatial diffusion. At lower energy, the gyroradius is small and the diffusion process can settle within a distance smaller than $\lambda_{\mathrm{c}}$. As resonant interactions occur with magnetic modes having wavelengths $\lambda\sim r_{\mathrm{g}} < \lambda_{\mathrm{c}}$, the particles start to diffuse in a region (or on a scale) where the average field does not vanish. A one-dimensional diffusion then settles along the residual mean field, with $D_{\parallel} > D_{\perp}$, and the full 3D diffusion is reached on larger timescales, when particles explore different uncorrelated cells of magnetic field. This causes the bumps seen on Fig.~\ref{fig:IEDCs}b.

\begin{figure*}[t]
\hfill
\includegraphics[width=0.47\linewidth]{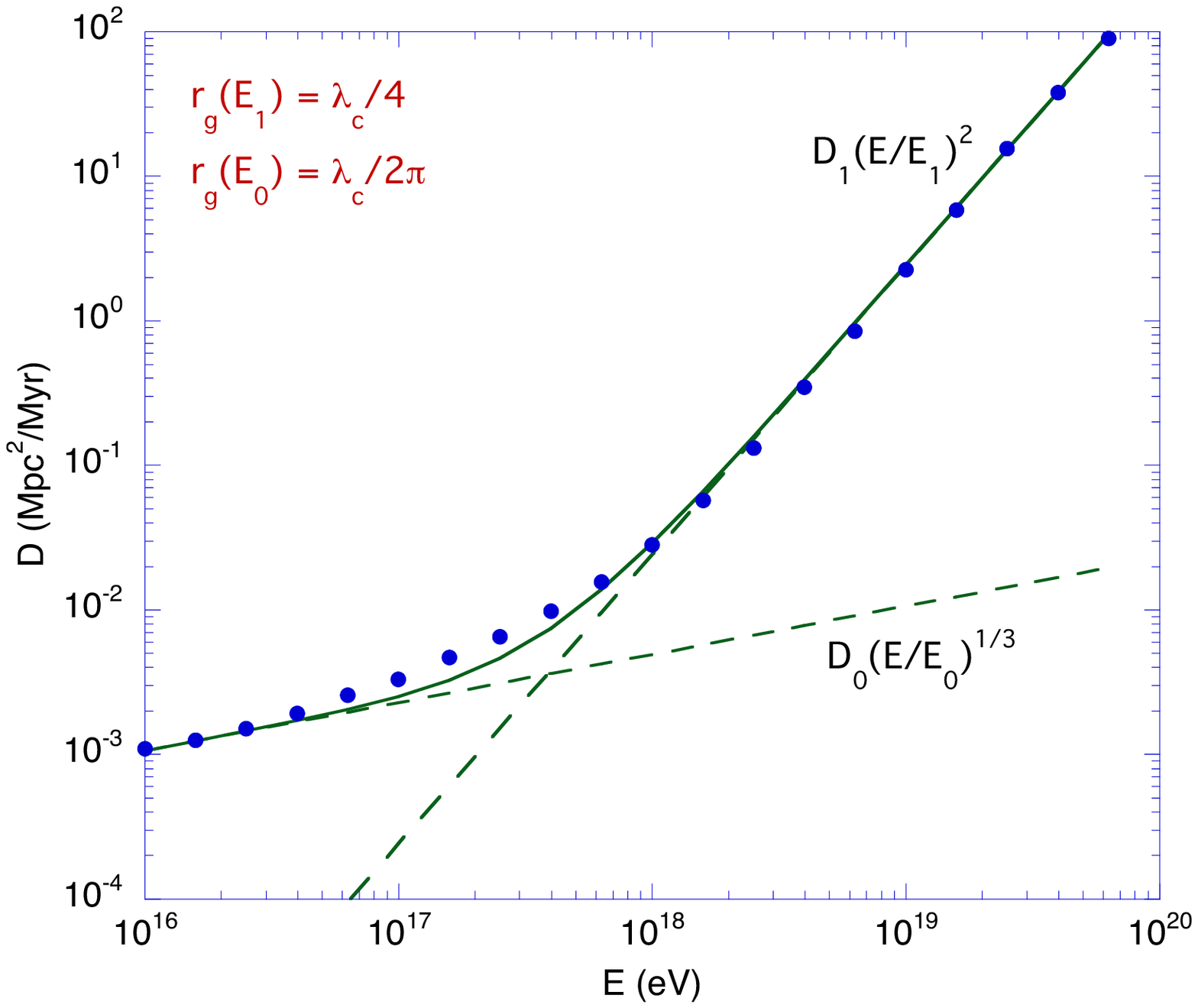}
\hfill
\includegraphics[width=0.47\linewidth]{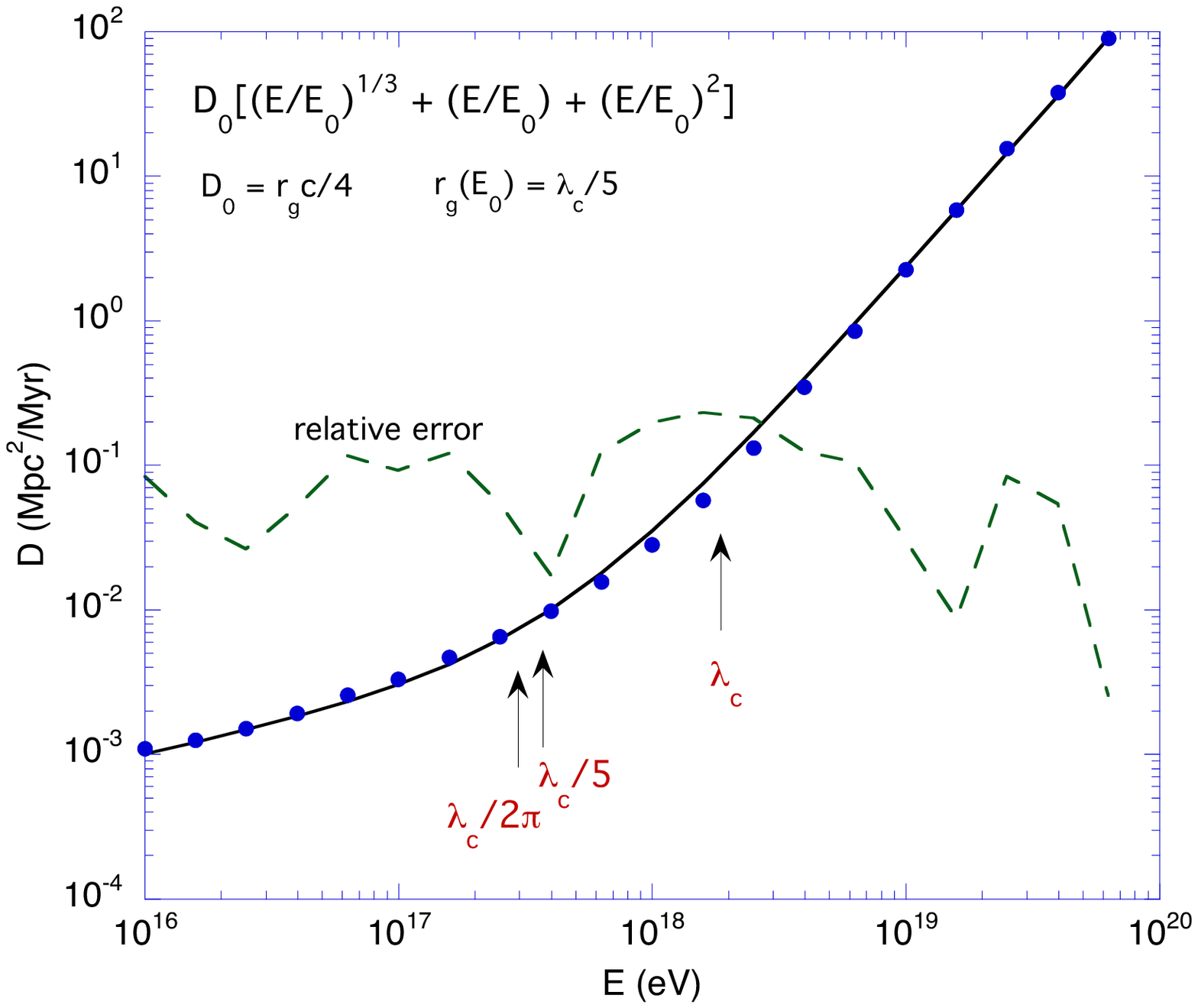}
\hfill
\vspace{-1cm}
\caption{Diffusion coefficient as a function of energy ($B = 10$~nG, $\lambda_{\mathrm{max}} = 1$~Mpc), with power-law fits (see text). On the right, the accuracy of the fit is also shown.}
\label{fig:DiffCoeff}
\end{figure*}

Quantitatively, the diffusion coefficients are given by the values of the plateau in Fig.~\ref{fig:IEDCs}. They are gathered in Fig.~\ref{fig:DiffCoeff}, showing the smooth transition between the \emph{quasilinear regime}, where $D(E)\propto E^{1/3}$, and the \emph{non-resonant regime}, where $D(E)\propto E^2$, in excellent agreement with theoretical expectations (e.g.~\cite{AloBer04}). The low-energy limit is well described by the approached formula $D(E) = D_{\mathrm{Bohm}}(E_{0})\times (E/E_{0})^{1/3}$, where $r_{\mathrm{g}}(E_{0}) = \lambda_{\mathrm{c}}/2\pi$, and the high-energy limit by $D(E) = D_{\mathrm{Bohm}}(E_{1})\times (E/E_{1})^{1/3}$, where $r_{\mathrm{g}}(E_{1}) = \lambda_{\mathrm{c}}/4$. The sum of these two functions provides a reasonable fit of $D(E)$ at all energies (see Fig.~\ref{fig:DiffCoeff}a).

An even better fit is obtained by adding a Bohm-like component, $D(E) = D_{\ast}(E/E_{\ast})$, and with only one critical energy, $E_{\ast}$:
\begin{equation}
D(E) = D_{\ast}\left[\left(\frac{E}{E_{\ast}}\right)^{1/3} + \left(\frac{E}{E_{\ast}}\right) + \left(\frac{E}{E_{\ast}}\right)^{2}\right],
\label{eq:DFit}
\end{equation}
where
\begin{equation}
r_{\mathrm{g}}(E_{\ast}) \equiv \frac{\lambda_{\mathrm{c}}}{5}\quad\mathrm{and}\quad D_{\ast}\equiv \frac{1}{4}cr_{\mathrm{g}}(E_{\ast}).
\label{eq:DStar}
\end{equation}
The accuracy is better than $\sim 10$\% (see Fig.~\ref{fig:DiffCoeff}b).

As can be seen, the diffusion coefficients can be obtained straightforwardly from one single parameter, $E_{\ast}$, which gathers the relevant information about the magnetic field intensity and coherence length.

\begin{figure}[t]
\hfill\includegraphics[width=0.9\linewidth]{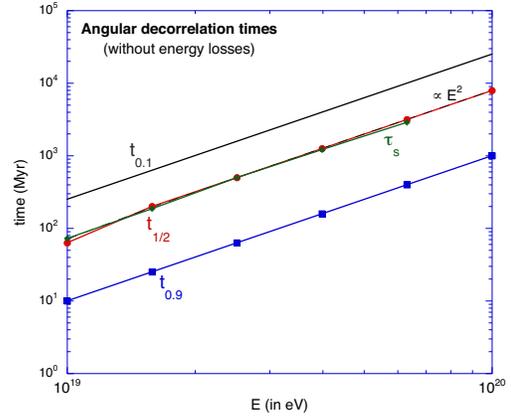}\hfill
\vspace{-1cm}
\caption{Decorrelation time and scattering time as a function of energy: $t_{0.9}$, $t_{1/2}$ and  $t_{0.1}$ are the time such that $\left<\cos(\theta(t) - \theta(0))\right> = 0.9$, 0.5 and 0.1, resp.; $\tau_{\mathrm{s}}\equiv 3D/c^2$ (see text).}
\label{fig:scatteringTime}
\end{figure}

\begin{figure*}[t]
\hfill
\includegraphics[width=0.43\linewidth]{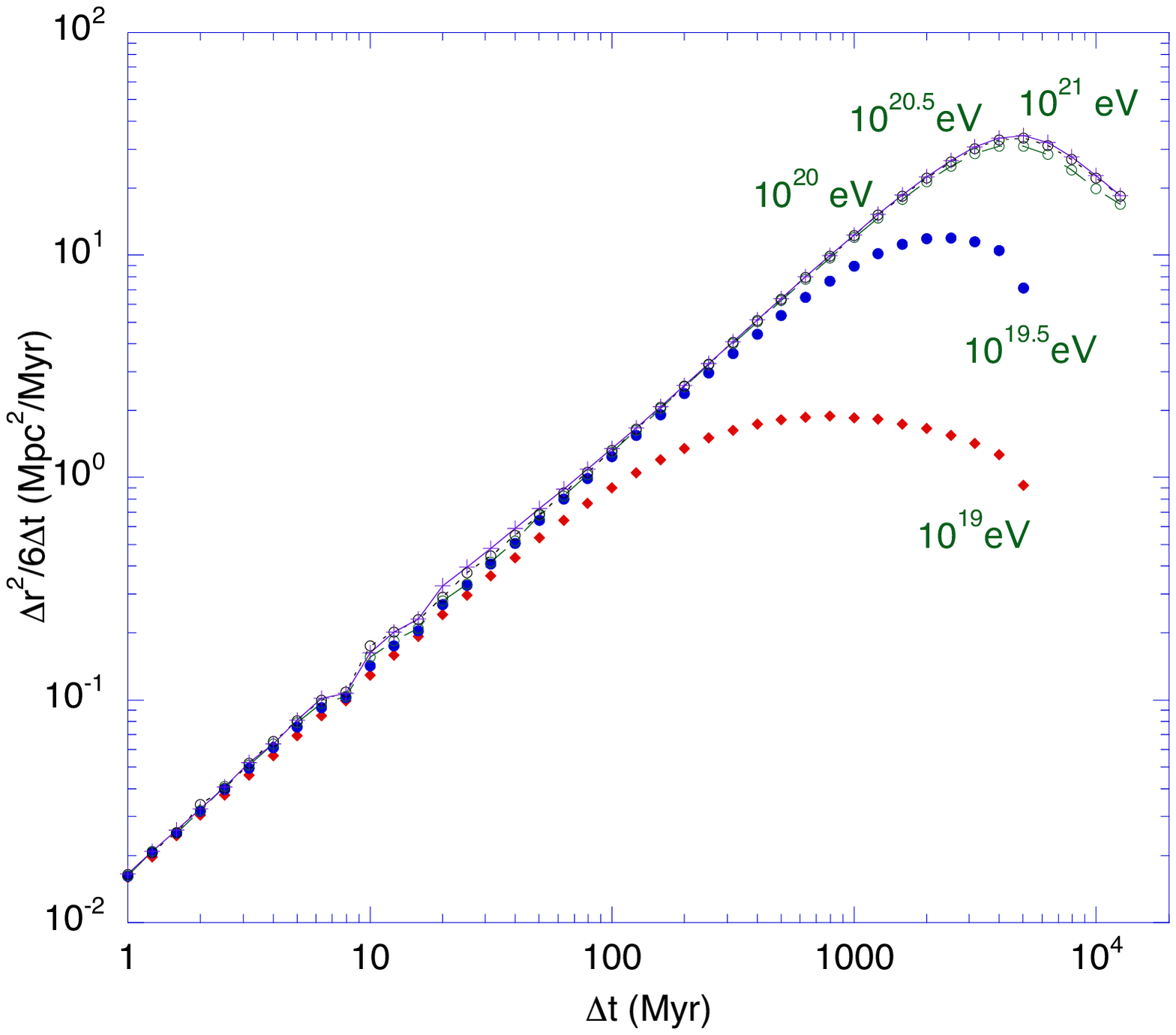}
\hfill
\includegraphics[width=0.43\linewidth]{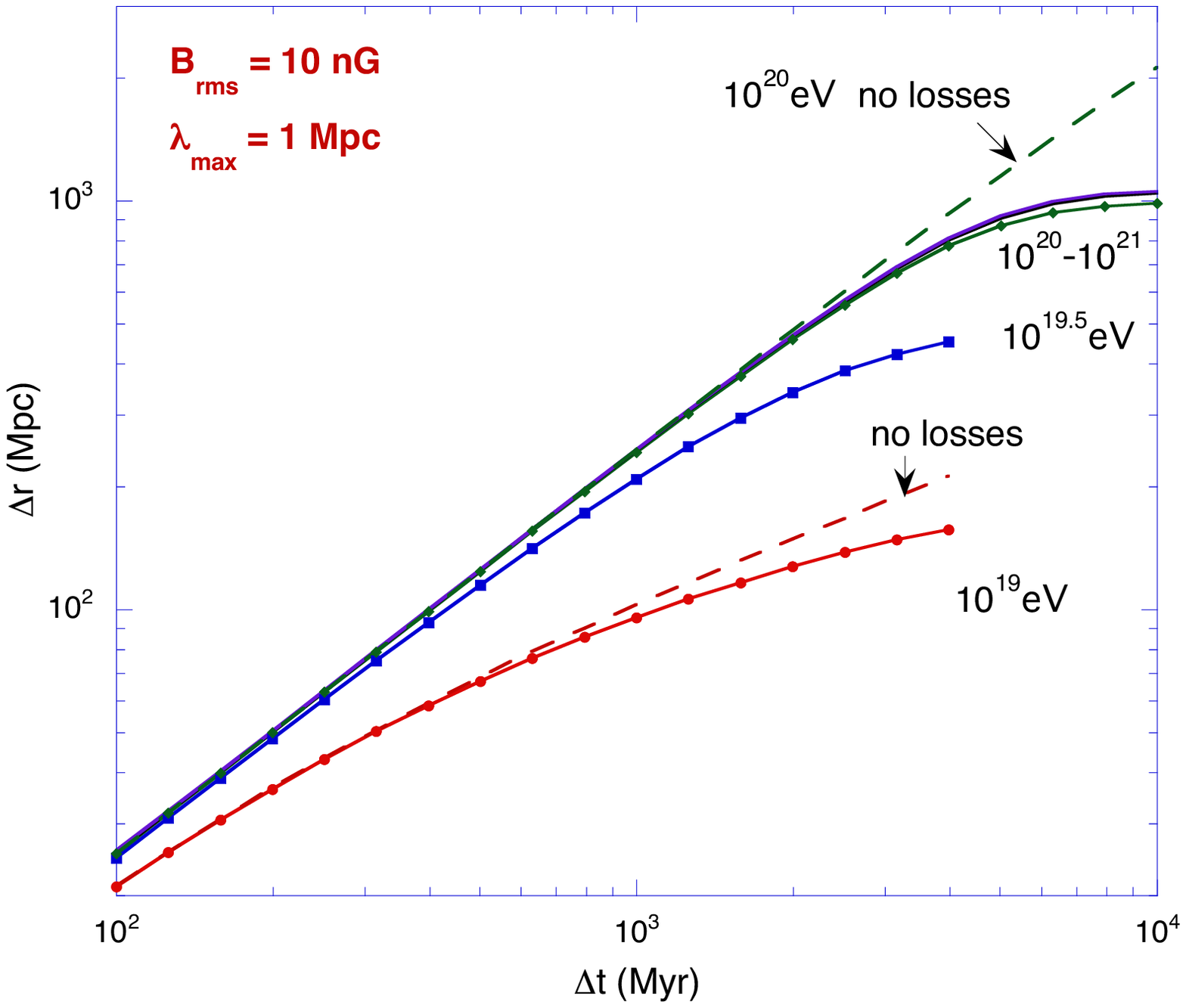}
\hfill
\vspace{-1cm}
\caption{Left: instantaneous effective diffusion coefficient (same as Fig.~\ref{fig:IEDCs}), taking into account proton energy losses. Right: Distance traveled by CRs of different energies, with and without energy losees, as a function of time.}
\label{fig:DiffCoeffWithLosses}
\end{figure*}

As the slope of $D(E)$ in logarithmic scales goes from 1/3 to 2, one may identify an approximate Bohm scaling (slope 1) for a limited range of energies~\cite{Casse+02}. However, even there the Bohm diffusion coefficient, $\frac{1}{3}r_{\mathrm{g}}c$, is never reached, the ``closest approach'' being for $E_{0}$, with $D \simeq 3D_{\mathrm{Bohm}}$. It should be noted here that while Casse et al.\cite{Casse+02} found $D(E)\propto E^{7/3}$ for $E \gg E_{0}$, our more precise calculations confirm the theoretical expectations in the non-resonant regime, namely $D(E)\propto E^2$. This scaling is also confirmed to be due to the $E^2$ dependence of the scattering time. In Fig.~\ref{fig:scatteringTime}, we plot the angular decorrelation times, $t_{0.9}$, $t_{1/2}$ and $t_{0.1}$, derived from Fig.~\ref{fig:angularCorrelation}, such that $\left<\cos\delta\theta(t)\right> = 0.9$, 0.5 and 0.1, respectively. Also plotted is the scattering time, defined from the random walk modeling of the diffusion process: $\tau_{\mathrm{s}} \equiv 3D/c^2$. As can be seen, $\tau_{1/2}$ gives an excellent definition of $\tau_{\mathrm{s}}$. An $E^2$ law is also shown to perfectly fit the curves.

\section{Transition from ballistic to diffusive regime}

As shown above, the propagation of CRs in a turbulent magnetic field smoothly passes from a ballistic regime, where the deflections are small and the distance traveled by the CRs away from their sources grows as $r\sim ct$, to a diffusive regime where it goes (on average) as $r\sim\sqrt{4Dt}$. In principle, the diffusion regime can always be reached, provided that one waits for a long enough time (a few $D/c^2$). To be specific, one may define $\tau_{\mathrm{diff}}(E)\equiv 4D(E)/c^2$ as the time required to reach the diffusion regime. Likewise, $\lambda_{\mathrm{diff}}(E)\equiv c\tau_{\mathrm{diff}} = \sqrt{4D\tau_{\mathrm{diff}}} = 4D(E)/c$ is the distance traveled away from the source before CRs of energy $E$ isotropize and diffuse.

However, CRs may not ``survive'' long enough to enter the diffusion regime. When energy losses are included in the propagation code, the IEDC curves look like in Fig.~\ref{fig:DiffCoeffWithLosses}a. A diffusion plateau is never reached, because the particles lose energy and their instantaneous diffusion coefficient then drops to lower and lower values. In Fig.~\ref{fig:DiffCoeffWithLosses}b, the traveled distance is shown to ``saturate'', as CRs travel a negligible distance at low energy. With the assumed field parameters, all particles above $10^{20}$~eV have the same transport properties, because they pass below the photo-pion production threshold while they are still in the ballistic regime.

\begin{figure*}[t]
\hfill
\includegraphics[width=0.45\linewidth]{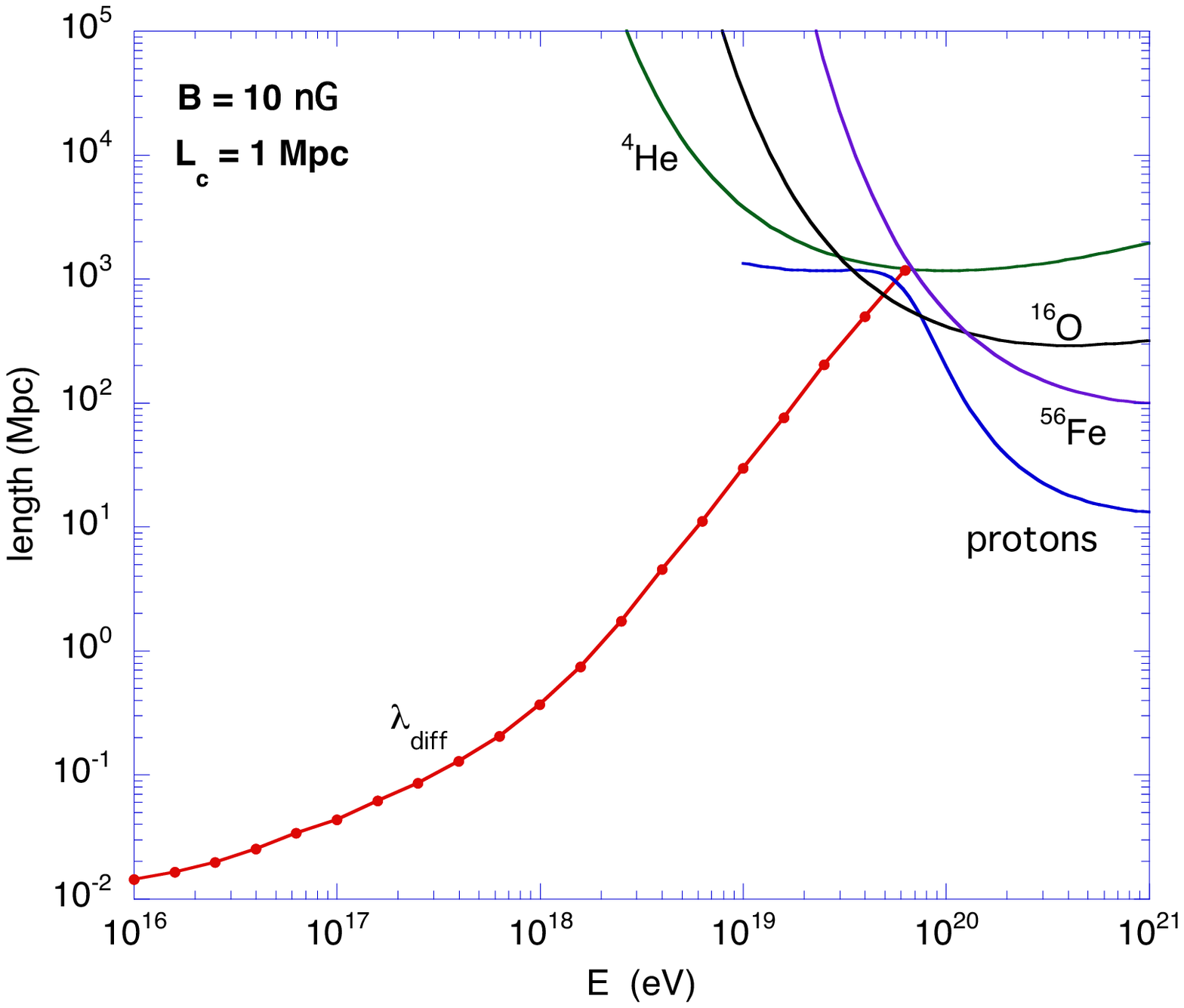}
\hfill
\includegraphics[width=0.45\linewidth]{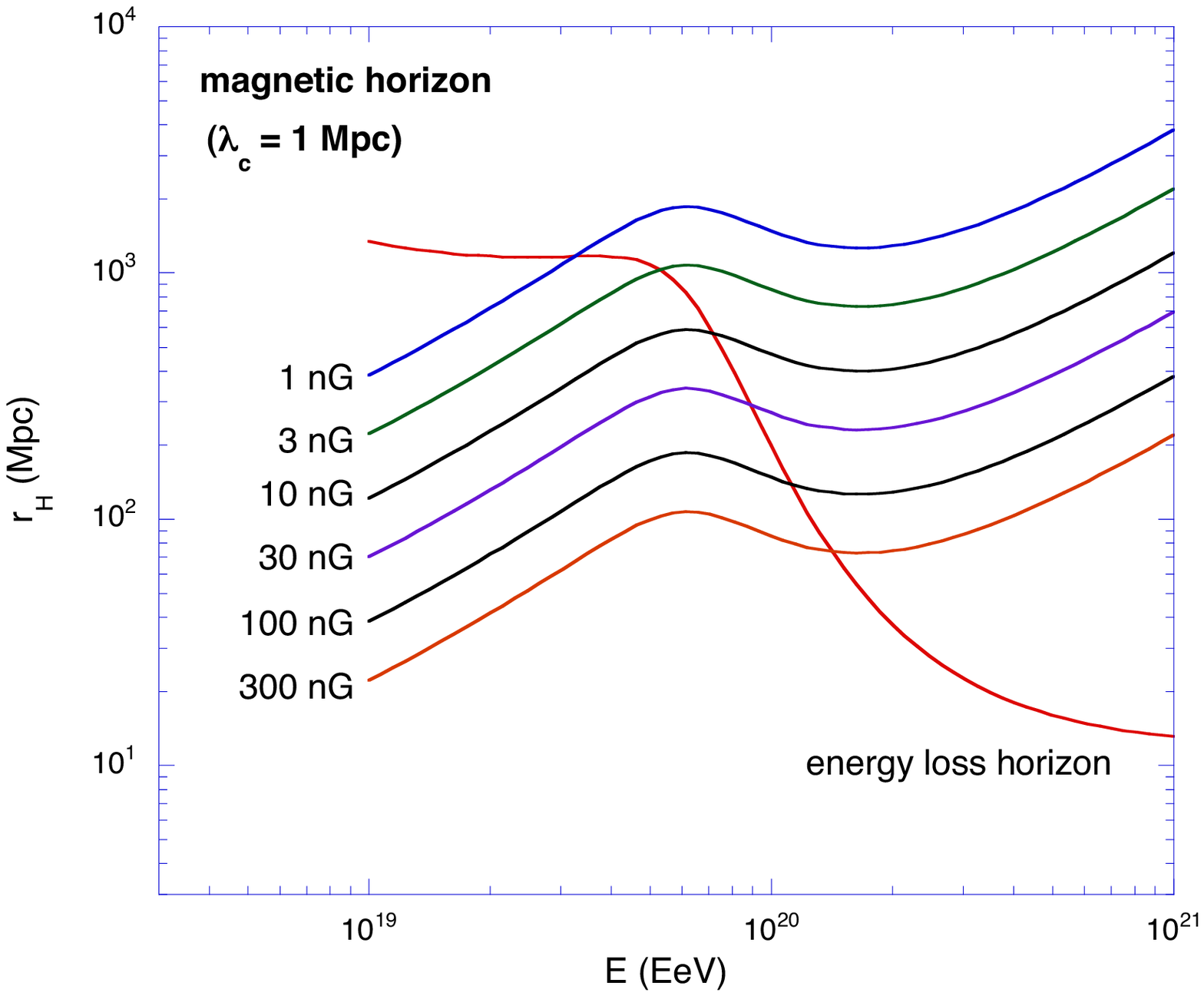}
\hfill
\vspace{-1cm}
\caption{Left: energy loss range and diffusion length, $\lambda_{\mathrm{diff}}=4D/c$, as a function of energy. Right: magnetic horizon, for various values of $B$.}
\label{fig:energyLossRange}
\end{figure*}

Interestingly, for the field parameters considered here, the transition between ballistic and diffusive regimes for particles in the GZK range (i.e. with energies between $3\,10^{19}$~eV and $10^{20}$~eV), occurs just around the \emph{GZK horizon}, $R_{\mathrm{GZK}}(E) = c\tau_{\mathrm{loss}}(E)$ (where the energy loss time scale is defined as $\tau_{\mathrm{loss}}(E)\equiv \dot{E}/E$). This is illustrated in Fig.~\ref{fig:energyLossRange}a, where we plot $R_{\mathrm{GZK}}(E)$ for protons, $^4$He, $^{16}$O and $^{56}$Fe nuclei, together with the above-defined diffusion length, $\lambda_{\mathrm{diff}}(E)$. This change of propagation regime in the GZK region of the CR spectrum is one of the ingredients of a possible alteration of the standard GZK feature, as we now discuss (see \cite{Deligny+04} for more details).

\section{Magnetic horizons and spectrum modification}

To see how magnetic fields can modify the appearance of the CR spectrum around the GZK feature, usually calculated with $B = 0$, it is useful to introduce the concept of \emph{magnetic horizon}, i.e. the maximum distance which (most) CRs can travel away from their source in a given magnetic field. This is set by the energy loss time, $\tau_{\mathrm{loss}}$ (or the age of the source, $t_{\mathrm{s}}$, if it is smaller), and the diffusion coefficient (or the IEDC if the diffusion regime is not reached). For high energy particles, roughly propagating in straight line, the magnetic field has no influence and one simply gets the usual GZK horizon. For low energy particles, the diffusion process prevents significant propagation beyond the magnetic horizon:
\begin{equation}
R_{\mathrm{magn}}(E) \simeq \sqrt{4D(E)\tau_{\mathrm{loss}}(E)}.
\label{eq:RMagn}
\end{equation}

According to the standard GZK argument, the CR flux above $10^{20}$~eV should be sharply reduced compared to that below $3\,10^{19}$~eV, say, because of the sudden decrease in the GZK horizon (cf. Fig.~\ref{fig:energyLossRange}a). However, in the presence of relatively high magnetic fields, low-energy particles \emph{also} have a limited range. In this sense, magnetic fields behave as ``low-cut filters'', whereas the CMB makes the universe a ``high-cut filter'' (GZK effect). The association and tuning of such filters (in series or in parallel) can lead to various shapes of the propagated CR spectrum, different from the simple, ``universal'' GZK feature.

In Fig.~\ref{fig:energyLossRange}b, we plot both the magnetic and GZK horizons for protons, as a function of energy. Of course, Eq.~(\ref{eq:RMagn}) does not apply when $\tau_{\mathrm{loss}} < \tau_{\mathrm{diff}}$, equivalent to $R_{\mathrm{magn}} > R_{\mathrm{GZK}}$, so that the actual CR horizon is always the smallest of the two values (lowest curve on Fig.~\ref{fig:energyLossRange}b). In the diffusion regime, one may rewrite the horizon radius as $r_{\mathrm{H}} = \lambda_{\mathrm{diff}}(\tau_{\mathrm{loss}}/\tau_{\mathrm{diff}})^{1/2}$ or numerically:
\begin{equation}
r_{\mathrm{H}}\simeq 0.58\,\mathrm{Mpc}\,\frac{E_{\mathrm{EeV}}}{Z B_{\mathrm{nG}}^\frac{1}{2}} \left(\frac{\tau_{\mathrm{loss}}}{1\,\mathrm{Myr}}\right)^{\frac{1}{2}}\hspace{-5pt}
\left(\frac{\lambda_{\mathrm{c}}}{1\,\mathrm{Mpc}}\right)^{-\frac{3}{2}}
\label{eq:horizonRadiusNum}
\end{equation}

As can be seen on Fig.~\ref{fig:energyLossRange}b, in a 10~nG field, CRs of $10^{19}$~eV cannot come from sources more distant than 100~Mpc, i.e. a volume 1000 times smaller than in the $B \simeq 0$ case. The \emph{same number of sources} should thus contributes at $10^{19}$~eV and $10^{20}$~eV! This does not mean, however, that there should not be any GZK feature in that case. Indeed, while magnetic fields prevent CRs from diffusing far from their sources, they also increase the CR density around each source, i.e. within the magnetic horizon. This is nothing but the magnetic confinement effect, very familiar for Galactic CRs. Since the number of particles remains the same, both effects exactly compensate, provided that the \emph{magnetic confinement spheres} (of radius $r_{\mathrm{H}}$) centered on the different sources are big enough to merge. If this is the case, the propagated CR spectrum with and without magnetic field are identical: the spatial horizon of the original GZK argument simply translates into a \emph{time horizon}, CRs at $10^{19}$~eV being able to contribute for a much longer time to the observed flux than CRs above $10^{20}$~eV. In fact, the propagated spectrum only depends on the time-of-flight distribution of the detected CRs, and it obviously makes no difference whether the trajectories were curved or not (see also ref.~\cite{AloBer04}).

If the magnetic confinement spheres do \emph{not} fully merge, however, interesting spectral effects can appear. The exact shape of the UHECR spectrum thus depends crucially on the source granularity, i.e. the typical distance between most important sources, $\Delta R_{\mathrm{s}}$~\cite{Deligny+04}.

\begin{figure}[t]
\hfill
\includegraphics[width=0.40\linewidth]{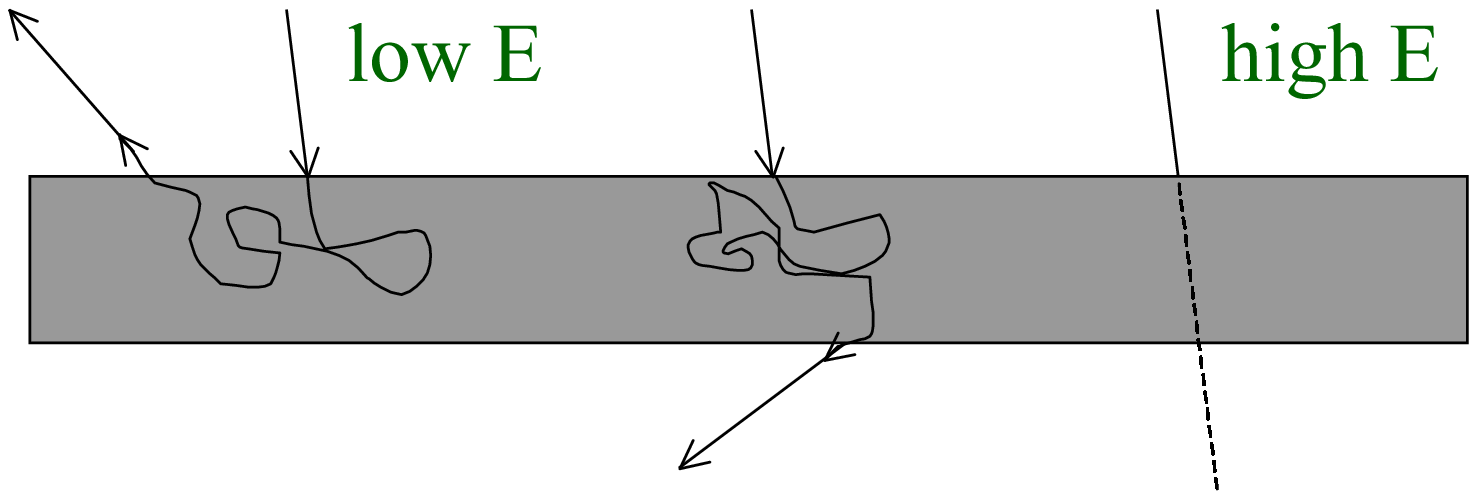}
\hfill
\includegraphics[width=0.40\linewidth]{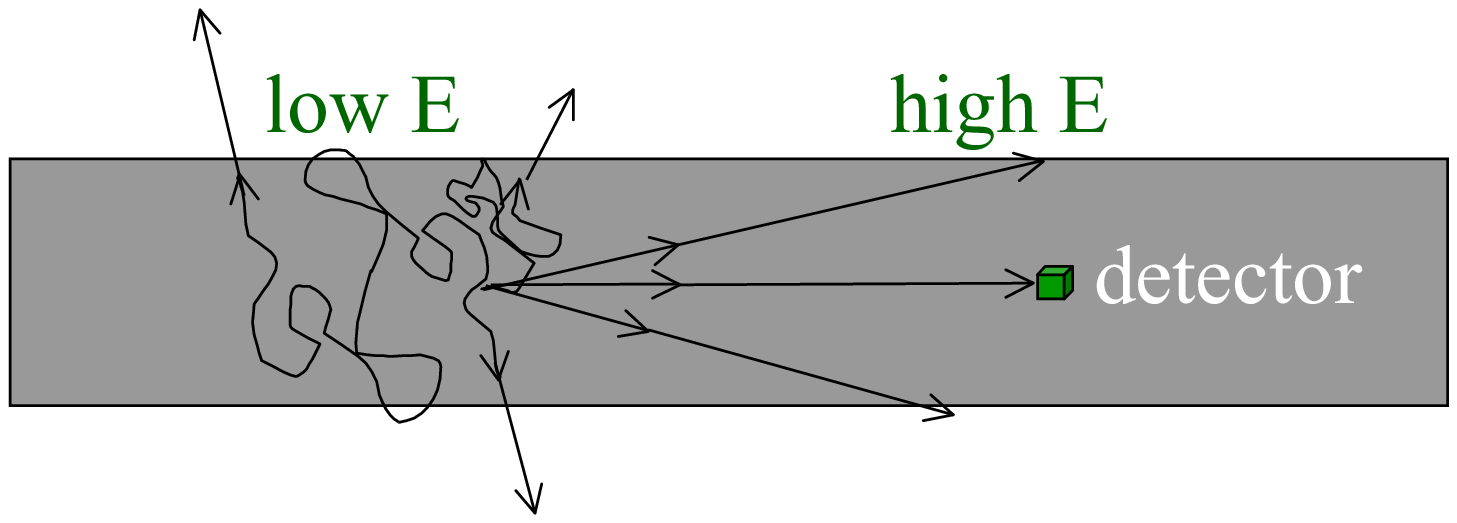}
\hfill
\vspace{-1cm}
\caption{Filtering of low-energy CRs by magnetized filaments (left) and out of the SGP (right).}
\label{fig:magneticFiltering}
\end{figure}

\begin{figure}[t]
\hfill
\includegraphics[width=0.91\linewidth]{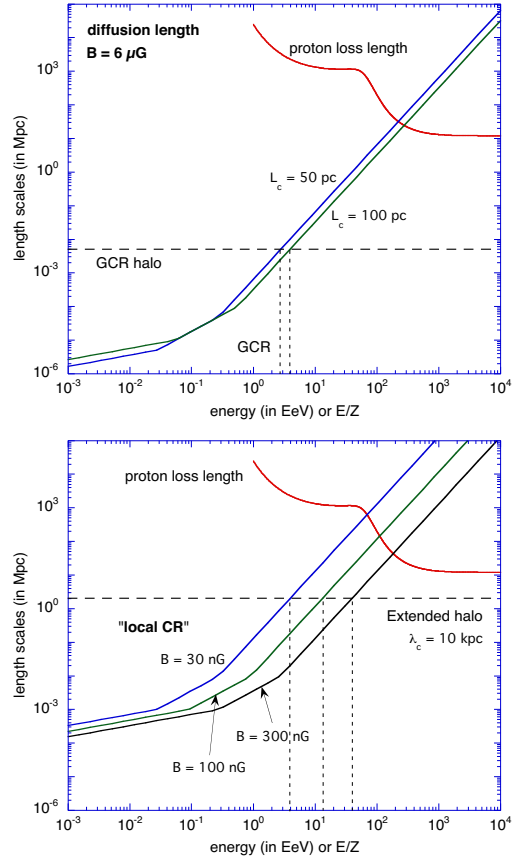}
\hfill
\vspace{-1cm}
\caption{CR confinement in magnetized structures. Top: GCR halo (5~kpc size), with different values of the field coherence length. Bottom: extended halo (assumed size of 2~Mpc, e.g. a joint superhalo with M31), with $\lambda_{\mathrm{c}} = 10$~kpc and different values of the superhalo magnetic field.}
\label{fig:magneticHalos}
\end{figure}

We simply mention here a few typical effects. First, if we are outside the confinement sphere around a source, it is clear that we cannot receive any CR from it. This may be of crucial importance to understand the transition between Galactic and extragalactic CRs. According to some models~\cite{Berezinsky+04}, this could occur around $3\,10^{17}$~eV. However, CRs of that energy have a relatively low diffusion coefficient, and thus a small magnetic horizon. In order for the source distance, $D_{\mathrm{s}}$, to be smaller than $r_{\mathrm{H}}$, the magnetic field (between the source and the detector) must be lower than (see Eq.~\ref{eq:horizonRadiusNum}):
\begin{equation}
B_{\mathrm{max}} \hspace{-2pt}\simeq\hspace{-2pt}
3.6\,\mathrm{nG}\left[\frac{D_{\mathrm{s}}}{5\,\mathrm{Mpc}}\right]^{\hspace{-2pt} -2} \hspace{-4pt} \left[\frac{E/Z}{3\,10^{17}\mathrm{eV}}\right]^{\hspace{-2pt}2} \hspace{-3pt} \left[\frac{\tau_{\mathrm{loss}}}{3\,\mathrm{Gyr}}\right]
\label{eq:BMax}
\end{equation}
which is roughly the ``equipartition field'' at that energy (Eq.~\ref{eq:BEq}). For such values, $3\,10^{17}$~eV CRs cannot come from sources more distant than 5~Mpc (or 3~Mpc if $B = 10$~nG), which leaves very few potential sources. This condition can be made even stronger if one considers that the source and the detector are in much larger fields, namely a few $\mu G$ inside galaxies. From a general point of view, CRs first have to get out of their local environment, then diffuse towards our Galaxy, and finally reach the Earth -- all within $\tau_{\mathrm{loss}}$ (or $t_{\mathrm{s}}$ if it is smaller). Trajectories in high field regions are longer, which tends to suppress the low-energy CRs (low-cut filter).

Other interesting situations may arise from sources located behind a filament or a galaxy cluster with high magnetic field. As illustrated on Fig.~\ref{fig:magneticFiltering}a, high energy particles are not affected, while low-energy CRs get a much longer path (or can even be reflected), which increases their energy losses. This will smoothen the GZK suppression, and in some cases prevent the lowest energy particles to reach us at all. By observing the CR spectrum coming from behind a highly magnetized filament or cluster with sufficient statistics, one should be able to observe a \emph{magnetic shadow} (in the form of a modified spectrum).

An other interesting effect appears if most sources \emph{and} our Galaxy are inside a \emph{contiguous} region of relatively high magnetic field (e.g. the supergalactic plane, SGP~\cite{SGP}). Low-energy CRs then diffuse out of the SGP on a scale comparable to its thickness, while high-energy CRs can reach us (cf. Fig.~\ref{fig:magneticFiltering}b). Magnetic fields thus again behave as low-cut filters, modifying the propagated spectrum.

Finally, we note that the magnetic confinement of GCRs generally applies to particles with a diffusion length smaller than the height scale of the so-called cosmic-ray halo, inferred from low-energy CR phenomenology to be of the order of $\sim 5$~kpc\cite{StrMos01}. It is shown in Fig.~\ref{fig:magneticHalos}a that this corresponds to CRs of energy lower than $\sim 3\,10^{18}$~eV, i.e. up to the ankle of the CR spectrum. However, one may consider an \emph{extended halo} with a much smaller magnetic field but a larger scale, where CRs of higher energy could still be confined, although with a smaller enhancement factor. In Fig.~\ref{fig:magneticHalos}b, we show that a magnetic field of 300~nG could confine protons in an extended halo of $\sim 2$~Mpc up to GZK energies, which means that Galactic sources might be relevant even for the highest part of the spectrum (of course, collimated or intermittent sources, such as GRBs, would then be required, in order to avoid a direct irradiation of the Earth, which would violate current limits on the CR anisotropy at these energies). This is all the more true if the most energetic CRs are heavy nuclei. An extended halo with a field of a few tens of nG is then sufficient to confine UHECRs injected by Galactic sources.

In conclusion, the study of the UHECR spectrum should involve a discussion of the magnetic field inside and outside galaxies.

\section*{Acknowledgments}
I wish to thank Denis Allard for his important help on the simulation of the diffusion coefficients in a turbulent field.

\end{document}